\documentclass[
    ,final            
  ]
  {aipproc}

\layoutstyle{6x9}

\def\approxgt{\mathrel{\hbox{\rlap{\lower.55ex \hbox {$\sim$}}
        \kern-.3em \raise.4ex \hbox{$>$}}}}
\def\approxlt{\mathrel{\hbox{\rlap{\lower.55ex \hbox {$\sim$}}
        \kern-.3em \raise.4ex \hbox{$<$}}}}
\def \xmm {\emph{XMM-Newton} }

\def\ltsima{$\; \buildrel < \over \sim \;$}
\def\lsim{\lower.5ex\hbox{\ltsima}}
\def\gtsima{$\; \buildrel > \over \sim \;$}
\def\gsim{\lower.5ex\hbox{\gtsima}}
\def\msole{~M_{\odot}}
\def\msun{~M_{\odot}}

\def\Msole{~M_{\odot}}
\def\Rsole{~R_{\odot}}
\def\Mdot {\dot M}
\def\hd {HD\,49798}
\def\rx {RX\,J0648.0--4418}

\def\hr {HD\,49798/RX\,J0648.0--4418}

\newcommand{\apj}{{\it ApJ}}

\newcommand{\apjl}{{\it ApJ}}
\newcommand{\aap}{{\it A\&A}}

\newcommand{\mnras}{{\it MNRAS}}

\newcommand{\physrep}{{\it Physics Reports}}

\begin{document}

\title{The discovery of a massive white dwarf in the peculiar
binary system HD 49798/RX J0648.0--4418}


\author{S. Mereghetti}{
  address={INAF IASF-Milano, v. E.Bassini 15, 20133 Milano, Italy}
}

\author{A. Tiengo}{
  address={INAF IASF-Milano, v. E.Bassini 15, 20133 Milano, Italy}
}

\author{P. Esposito}{
  address={INAF IASF-Milano, v. E.Bassini 15, 20133 Milano, Italy}
  altaddress={INFN-Pavia, v. A.Bassi 6, 27100 Pavia, Italy}
}

\author{N. La Palombara}{
  address={INAF IASF-Milano, v. E.Bassini 15, 20133 Milano, Italy}
}

\author{G. L. Israel}{
  address={INAF Osservatorio Astronomico di Roma, v. Frascati 33, 00040 Monteporzio Catone, Italy}
}

\author{L. Stella}{
  address={INAF Osservatorio Astronomico di Roma, v. Frascati 33, 00040 Monteporzio Catone, Italy}
}

\begin{abstract}

An \xmm\ observation performed in  May 2008 has confirmed that the 13 seconds
pulsations in the X-ray binary \hr\ are due to a rapidly rotating white dwarf.
From the pulse time delays induced by the 1.55 days orbital motion, and
the system's inclination, constrained by the duration of the X-ray eclipse
discovered in this observation,
we could derive a mass of 1.28$\pm$0.05 $\msole$ for the white dwarf.
The future evolution of this post common envelope binary system will likely involve
a new phase of mass accretion through Roche-lobe overflow  that could drive
the already massive white dwarf above the Chandrasekhar limit and produce a Type Ia
supernova.

\end{abstract}

\maketitle


\section{Introduction}

White dwarfs have masses in a narrow range centered at
about 0.6 $\Msole$ (see, e.g., \cite{kep07}), but a few examples of massive white dwarfs
($>$1.2 $\Msole$) have been reported \cite{dah04,ven08}.
Massive white dwarfs in binary systems, where mass transfer can occur, are particularly
interesting since they are good candidates for the formation of Type Ia supernovae.

Using data obtained with the \xmm\ satellite, we have recently found compelling evidence
\cite{mer09} for the presence of a  white dwarf
with mass $>$1.2 $\Msole$ in the X--ray binary system \hr .
In the next section we describe the properties of this peculiar binary, composed
of a 13 s X--ray pulsar orbiting a hot subdwarf star. We then briefly review
the new \xmm\ results and discuss some of their implications.

\section{The peculiar binary \hr\ }

\hr\   is the only known binary system composed of
an early type subdwarf star and a pulsating X-ray source. The
optical/UV emission from  this system is dominated by the  bright star \hd :
with a V band magnitude of 8.3, this is the brightest known hot subdwarf.
Extensive observations carried out since the
time of its discovery led to classify it as a star of  sdO5.5 type,
and showed that \hd\ is a  single-lined spectroscopic binary with orbital period
P$_{ORB}$ =1.55 days \cite{tha70,kud78}.
The optical mass function could be measured with great precision,
f$_{OPT}$ = 0.263 $\pm$ 0.004 $\msun$  \cite{sti94},
but the nature of the companion star (unseen in the optical) remained obscure for decades.

The situation changed when the \textit{ROSAT} satellite detected a
strong flux of soft X-rays from this system and  periodic
X-ray pulsations at P=13.2 s were discovered  \cite{isr97}. This finding  showed
that the companion star
is either a neutron star or a white dwarf. However, as discussed
in \cite{bis97} and confirmed by our new data \cite{mer09}, the X-ray luminosity is much
smaller than that expected from a neutron star accreting in the
stellar wind of \hd\ \cite{ham81}, leading to the conclusion that \rx\ is a white dwarf.

 \begin{figure}
   \includegraphics[height=.4\textheight]{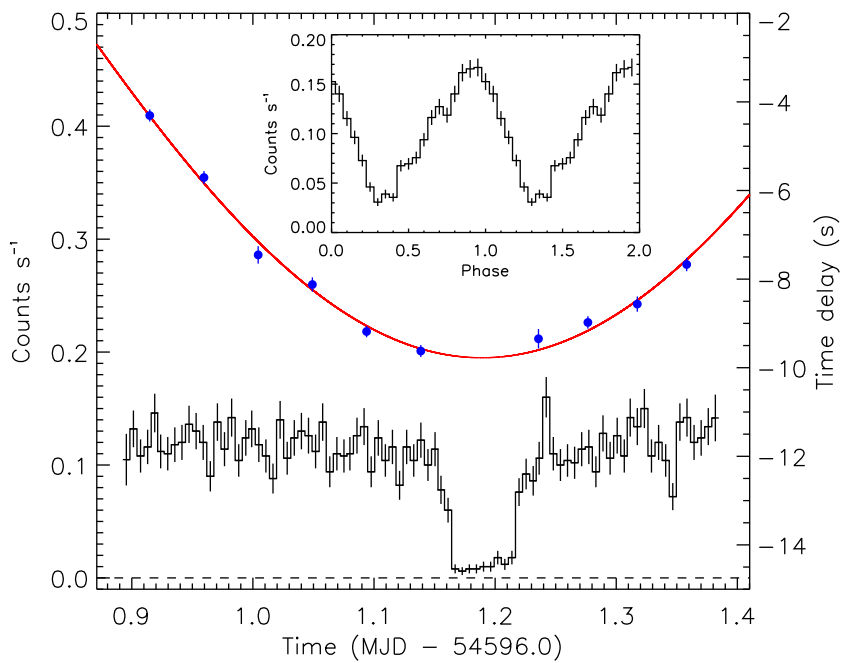}
   \caption{Black solid line (left axis): Light curve of \rx\ in the 0.15--0.4
keV  range showing an eclipse
lasting ~4700 s. The dashed line indicates the background level.
Blue points (right axis):  time delays of the
pulsations measured at different orbital phases  with the best fit sinusoid (red line).
Inset:  spin-period pulse profile in the 0.15--0.4
keV  (P=13.18425$\pm$0.00004 s).}
 \end{figure}

 \begin{figure}
   \includegraphics[height=.4\textheight,angle=-90]{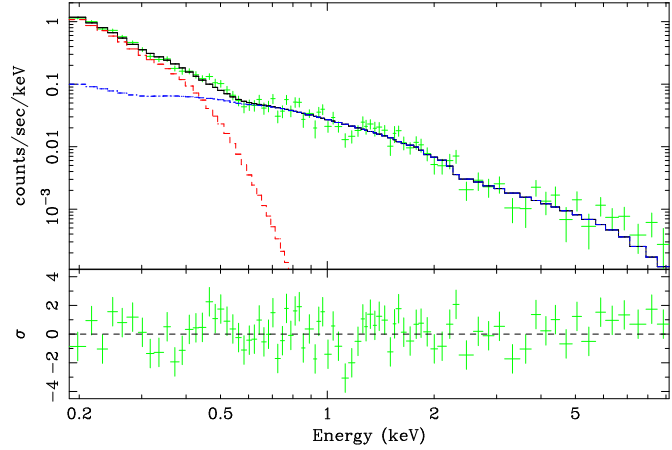}
   \caption{X--ray spectrum of \rx\ as measured with the EPIC pn camera on \xmm.
   Top panel: data points and best fitting  model.
   Bottom panel:  residuals from the model in units of standard deviations.
      }
 \end{figure}

\section{\xmm\ results }

The presence of fast X-ray pulsations makes this system equivalent
to a double spectroscopic binary, with the possibility to derive
all the orbital parameters  and, in particular, the masses of the two components.
With this objective, we observed \hr\ for 44 ks on May 10-11, 2008 with \xmm\.
Although four \xmm\ observations were already performed in 2002,
they are not particularly useful due to their very short duration \cite{tie04}.
Based on the well known optical ephemeris \cite{sti94},
the new observation was scheduled in such a way to include the orbital
phase at which an X-ray eclipse could be expected
($\Phi$=0.75), which  previous X-ray observations did not cover.
The 0.15--0.4 keV light curve (Fig.1), obtained with the EPIC instrument, clearly
shows the presence of an eclipse lasting $\sim$1.3 hours.
Since the radius of \hd\ is known from optical observations (R$_{C}$=1.45$\pm$0.25
$\Rsole$  \cite{kud78}),  the system inclination can be constrained
from the duration of the eclipse using the relation
(R$_{C}$/a)$^{2}$ = cos$^{2}$i + sin$^{2}$i sin$^{2}$$\Theta$, where a
is the orbital separation and $\Theta$ is the eclipse half angle.
This gives an inclination in the range  79-84$^{\circ}$.

From a timing analysis of the 13.2 s pulsations in the EPIC  data it was also possible
to measure the delays in the times of arrival of the pulses
induced by the orbital motion (Fig. 1) and thus obtain the
projected semi-major axis A$_{X}$ sin i = 9.78$\pm$0.06 light-s.
This corresponds to an X-ray mass
function f$_{X}$ =
0.419$\pm$0.008 $\Msole$ that, combined with the
optical mass function \cite{sti94} and with the inclination derived above,
yields the masses of the two stars:
M$_{X}$=1.28$\pm$0.05 $\Msole$ for the
white dwarf and M$_{C}$ = 1.50 $\pm$ 0.05 $\Msole$ for \hd .
Independently of the radius of \hd , a firm lower limit of
1.2 $\Msole$ (2$\sigma$ c.l.) on the white dwarf mass is obtained
for   i= 90$^{\circ}$.

The X-ray emission from \rx\ is very soft. As  shown in Fig.2
its spectrum can be fitted with
the sum of a blackbody with
temperature kT = 39 eV, contributing most of the observed flux,
and a power law with photon index $\sim$2.
For the distance of 650 pc, derived from optical observations \cite{kud78},
the X-ray luminosity in the 0.2-10 keV energy
range is $\sim$2$\times$10$^{31}$  erg s$^{-1}$.

\section{Discussion}

The above findings indicate that \rx\ is one of the most massive
white dwarfs currently known  and the one with the shortest spin period.
We note that most white dwarf masses are derived with
indirect methods, such as surface gravity estimates obtained by
spectral modelling, or the measurement of gravitational redshift.
These methods, beside depending on models and assumptions,
provide combined information on mass and radius, hence the
resulting masses cannot be used to determine independently the
mass-radius relation. Our mass determination for \rx\ is directly based  on a
dynamical measurement, thus making this star an ideal
target to better constrain the white dwarfs equation of state.
The mass lower limit, coupled  with the 13 s spin
period,  allows us to set an upper limit of 6000 km on the white
dwarf's radius, based only on simple rotational stability
considerations \cite{cha87}.

The high mass and fast spin of \rx\ likely result from a
previous evolutionary phase in which the accretion of mass and
angular momentum took place at a much larger rate than currently observed.
The accretion of mass down to the white dwarf surface implies that the
magnetic dipole is \ltsima 10$^{29}$
($\Mdot$ / 10$^{-8}$ $\Msole$ yr$^{-1}$)$^{15/16}$ (1350 km
s$^{-1}$ / V$_{WIND}$)$^{15/4}$ Gauss cm$^{3}$,  where we have normalized the
wind mass loss $\Mdot$ and terminal velocity V$_{WIND}$ to the values measured for \hd\ \cite{ham81}. A stronger magnetic field would prevent accretion through the
onset of the propeller effect.
Thus, \hr\ can be seen as  a white dwarf analogous of
low mass X-ray binaries in which weakly magnetic neutron stars are
spun-up to short spin periods and shine as recycled millisecond
radio pulsars when accretion onto them stops \cite{bha91}.

The future evolution of \hd\ will probably lead to a new phase of unstable mass transfer
through Roche-lobe overflow \cite{ibe94}, during which the
accretion of helium-rich material  might push the already massive white dwarf
above the  Chandrasekhar limit and trigger a Type Ia supernova explosion.


\begin{theacknowledgments}
  We acknowledge partial support from contract ASI/INAF I/088/06/0
\end{theacknowledgments}



\bibliographystyle{aipproc}   





\end{document}